\documentclass[a4paper,11pt]{article}
\usepackage{jheppub} % for details on the use of the package, please see the JINST-author-manual
\usepackage{lineno}
\usepackage{color}
\usepackage{graphicx}% Include figure files
\usepackage{dcolumn}% Align table columns on decimal point
\usepackage{bm}
\usepackage{amsthm}
\usepackage{tikz}
\usepackage{slashed}

\arxivnumber{2607.11634} % if you have one

\title{\boldmath Thermal Dilepton Polarization under Rotation or Magnetic Field in Heavy-ion Collisions}

\author{Haowen He,}
\author{Bowen Hou,}
\author{Xinyang Wang}
\emailAdd{wangxy@aust.edu.cn}
\author[1]{and Minghua Wei\note{Corresponding author.}}
\emailAdd{weiminghua@aust.edu.cn}
\affiliation{
Center for Fundamental Physics, School of Mechanics and Optoelectronic Physics,
Anhui University of Science and Technology,\\ Huainan, Anhui 232001, China }

\abstract{Dilepton (Virtual photon) polarization is characterized by anisotropic coefficients $\lambda_{\theta}$, $\lambda_{\phi}$, and $\lambda_{\theta\phi}$, which are expected to be influenced by vorticity and magnetic fields. This work investigates thermal dilepton production in a quark-gluon plasma via the quark-antiquark annihilation process $q\bar{q} \to \gamma^* \to l^+l^-$. Virtual photon polarization can be induced by both the spin polarization of quarks and the anisotropy of their momentum distribution in the medium. By employing the modified quark propagator under an external field, we derive the electromagnetic spectral function in a hot medium. Based on the spin-projection decomposition of the spectral function, the spin density matrix elements of the virtual photon and the anisotropy coefficients for the emitted dileptons are determined. Due to the distinct effects of vorticity and magnetic fields on the quark propagator, the resulting invariant mass spectra of dilepton polarization exhibit characteristic differences. Furthermore, our study reveals the response of dilepton polarization signals to external fields of varying strengths, suggesting dilepton polarization as a complementary and sensitive probe for both vorticity and magnetic fields in relativistic heavy-ion collisions.}

\begin{document}
\maketitle
\flushbottom
\section{INTRODUCTION}
Relativistic heavy-ion collisions (HICs) produce a new deconfined state of quantum chromodynamics (QCD) matter known as the quark-gluon plasma (QGP). Due to the immense orbital angular momentum generated in non-central collisions, the produced QGP exhibits the most vortical fluid behavior among all known physical systems, while simultaneously being immersed in a strong background electromagnetic field. The average vorticity can reach a magnitude of \(\Omega\sim10\) MeV~\cite{STAR:2017ckg}, with local fluctuations potentially yielding even higher values~\cite{Jiang:2016woz}. In the early stages of the collision, the magnetic field strength is estimated to be on the order of $eB\sim 1-10m_{\pi}^{2}$~\cite{Skokov:2009qp,Voronyuk:2011jd,Deng:2012pc}, where \(m_{\pi }\) denotes the pion mass. Since both the vorticity and magnetic field scales approach the typical QCD scale, QCD matter under these extreme conditions exhibits a rich variety of novel, non-trivial phenomena~\cite{Becattini:2021wqt,Kharzeev:2007jp,Chen:2015hfc,Ebihara:2016fwa,Liu:2017spl,Zhang:2018ome,Ambrus:2019ayb,Bai:2025rwx}.

The hot QGP emits electromagnetic probes, such as real photons and dileptons, which serve as direct messengers reflecting the thermodynamic and transport properties of the QCD medium~\cite{Rapp:2013nxa,Bailhache:2025kwa}. Once produced, these probes escape the interaction region without participating in the strong interaction with the surrounding medium, thereby offering an unperturbed window into the early, hot stages of the plasma. In this work, we focus specifically on thermal dileptons produced via the annihilation of quark-antiquark pairs within the QGP medium.

For decades, the invariant mass spectrum of dileptons has been utilized as a reliable thermometer for extracting the QGP temperature~\cite{Rapp:2014hha,Churchill:2023vpt,Churchill:2023zkk,Massen:2024pnj}, a utility robustly demonstrated by a wealth of recent experimental data~\cite{STAR:2024bpc}. Concurrently, dilepton physics has advanced considerably, with numerous theoretical studies devoted to exploring the influences of external magnetic and vorticity fields on the dilepton production rate~\cite{Bandyopadhyay:2016fyd,Ghosh:2020xwp, Chaudhuri:2021skc,Mondal:2023vzx,Bandyopadhyay:2025evc,Castorina:2007eb, Wei:2021dib,Heinz:2000bk,Castano-Yepes:2025zae}. These investigations have revealed that dilepton production exhibits a substantial elliptic flow coefficient \(v_{2}\) in the presence of strong vorticity or magnetic fields~\cite{Wei:2021dib,Wang:2022jxx, Das:2021fma,Salabura:2020tou,Mondal:2023ypq}. Owing to these developments, dileptons have emerged as a quantitative ``ruler'' for characterizing the vorticity and magnetic field strengths in the QGP. Beyond the production rate and azimuthal anisotropy \(v_{2}\), the spin polarization of dileptons has recently attracted significant interest as a more sensitive, multi-dimensional observable.

In a hot QGP medium, quarks and antiquarks annihilate into virtual photons ($q \bar{q} \rightarrow \gamma^{*}$)~\cite{Laine:2013vma,Jackson:2019mop}, which subsequently decay into lepton pairs. In HICs, these quarks and antiquarks can be spin-polarized by the vortical or magnetized medium; alternatively, their momentum distributions can become inherently anisotropic due to external fields. Consequently, the emitted virtual photons—and hence the resulting dileptons—inherit a specific polarization, manifesting as an anisotropic angular distribution of the lepton pairs in the virtual photon rest frame. Dilepton polarization thus provides a unique and complementary avenue for probing the microscopic properties and collective dynamics of the QGP~\cite{Shuryak:2012nf,Baym:2017qxy,Speranza:2016tcg,Speranza:2018osi,Wei:2024lah,Seck:2023oyt,Wu:2024vyc,Gao:2026vxs}. For instance, dilepton polarization is expected to be highly sensitive to the spatial and momentum anisotropy of the plasma~\cite{Coquet:2023wjk}. On the experimental front, the NA60 and HADES collaborations have measured the angular distributions of dileptons to extract the polarization signatures originating from hot and dense nuclear matter~\cite{NA60:2008iqj,HADES:2011nqx}. These efforts have spurred deeper theoretical inquiries into how extreme external conditions influence dilepton polarization in heavy-ion collisions.

As a recent example, Ref.~\cite{Wei:2024lah} evaluated dilepton polarization under relatively weak magnetic field conditions (\(\vert{}eB\vert{} \ll m_{\pi}^{2}\)). They found that the polarization coefficients, such as \(\lambda _{\theta}\) and \(\lambda _{\phi}\), exhibit significant sensitivity to both the direction and strength of the magnetic field, thereby offering a promising observable for experimental field measurements. However, that study was grounded in the semi-classical Boltzmann equation for a magnetized medium and did not incorporate the full quantum-mechanical effects of quarks under external fields, such as Landau level quantization. Furthermore, dilepton polarization induced by a background vorticity field remains largely unexplored within a fully quantum field-theoretic framework.

In this work, we develop a comprehensive theoretical framework based on the finite-temperature quantum field theory spectral function~\cite{Weldon:1990iw}. This approach is well-suited for investigating arbitrarily strong vorticity and magnetic fields, as it intrinsically accounts for the underlying quantum behavior of quarks. In a rotating or magnetized medium, the Dirac equation and the corresponding fermion propagators are modified accordingly. From the imaginary part of the photon polarization tensor, we evaluate the electromagnetic spectral function~\cite{Wei:2021dib,Wang:2022jxx, Das:2021fma,Salabura:2020tou}. We then employ a structural tensor decomposition of the spectral function, separating it into a longitudinal component and several independent transverse components~\cite{Baym:2017qxy,Wei:2024lah,Wu:2024vyc}. Utilizing the established relationship between the spectral function and the dilepton production rate~\cite{Weldon:1990iw}, we systematically investigate dilepton production and its angular distribution in the presence of either a vorticity field or a magnetic field. Our results demonstrate that the anisotropy coefficients of thermal dileptons exhibit qualitatively different behaviors between rotating and magnetized media.
 
This manuscript is organized as follows. In Sec.~\ref{sec:theory}, we present the general theoretical formulation of dilepton polarization. In Secs.~\ref{sec:vorticity} and \ref{sec:magnetic-field}, we derive the electromagnetic spectral functions in the presence of a vorticity field and a magnetic field, respectively. In Secs.~\ref{sec:rotating} and \ref{sec:magnetized}, we present our numerical results and discuss the distinct polarization signatures that emerge in rotating and magnetized QCD media. Finally, a summary and concluding remarks are given in Sec.~\ref{sec:summary}.

\section{Theory of dilepton polarization}
\label{sec:theory}
\begin{figure*}[t]
    \centering
    \includegraphics[width=0.8\linewidth]{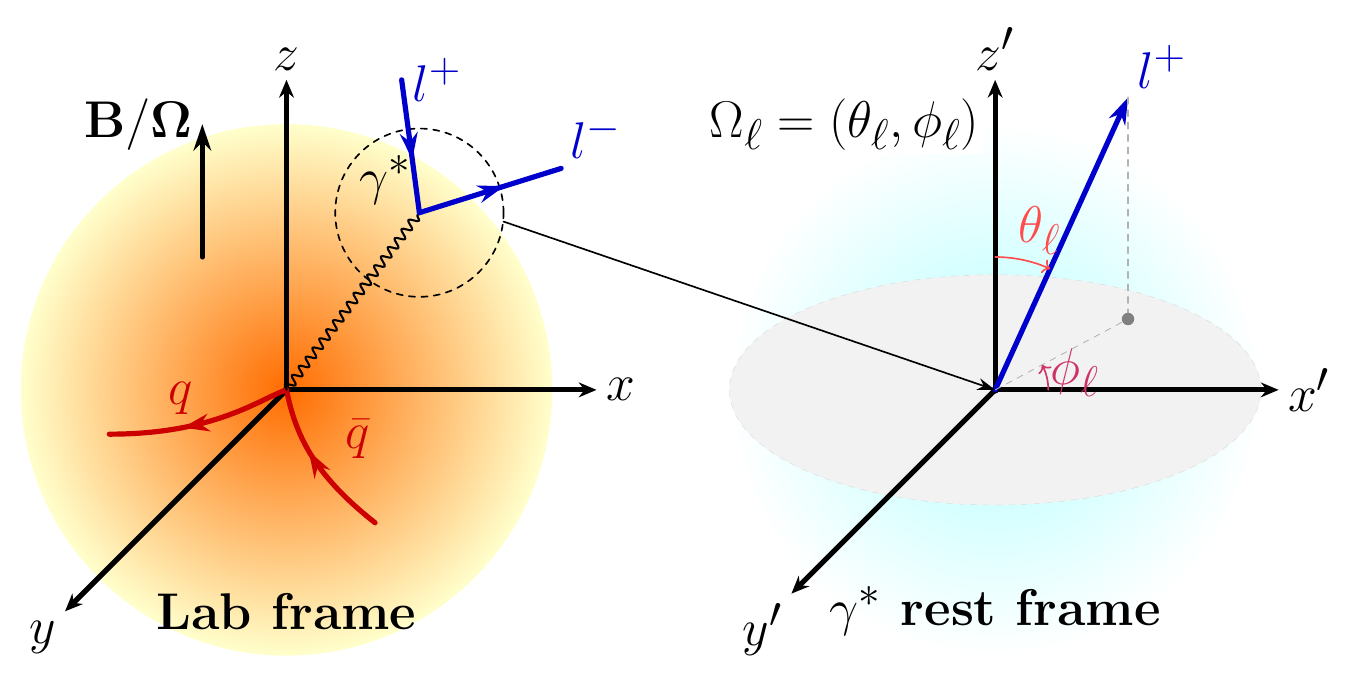}
    \caption{Schematic illustration of dilepton production and polarization in the presence of a vorticity field / magnetic field.}
    \label{fig:illustration}
\end{figure*}

In a hot QGP, quark-antiquark annihilation into a virtual photon $\gamma^{*}$, followed by its decay into a dilepton pair $l^{+}l^{-}$, constitutes the leading production mechanism under consideration. The left panel of Fig.~\ref{fig:illustration} provides a schematic depiction of this process. We define a lab frame with the $x$-axis along the impact parameter, the $y$-axis along the beam direction, and the $z$-axis perpendicular to the reaction plane $xOy$. The thermal dilepton production rate per unit volume is given by~\cite{McLerran:1984ay,Weldon:1990iw}
\begin{equation}
\label{eq:dR1}
    \frac{dR_{l^{+}l^{-}}}{d^{4}x} = 2\pi e^{2} L_{\mu\nu} \rho^{\mu\nu}(q) e^{-\beta \omega} \frac{\mathrm{d}^{3}\mathbf{l}_{1}}{(2\pi)^{3}E_{1}} \frac{\mathrm{d}^{3}\mathbf{l}_{2}}{(2\pi)^{3}E_{2}}\, ,
\end{equation}
where $l_{1}^\mu=(E_{1}, \mathbf{l}_{1})$ and $l_{2}^\mu=(E_{2}, \mathbf{l}_{2})$ denote the lepton and antilepton four-momenta, respectively, and $q=(\omega, \mathbf{q})$ is the virtual-photon four-momentum, with $\omega=E_{1}+E_{2}$ and $\mathbf{q}=\mathbf{l}_{1}+\mathbf{l}_{2}$. Here, $e$ is the elementary charge and $\beta=1/T$ the inverse temperature. The lepton tensor reads $L^{\mu\nu}=-4[(l_{1} \cdot l_{2} + m_l^2)g^{\mu\nu} - l_{1}^\mu l_{2}^\nu - l_{1}^\nu l_{2}^\mu]$, with $m_l$ being the lepton mass. The electromagnetic spectral function $\rho^{\mu\nu}(q)$ encodes the medium response and is sensitive to the presence of rotation or a magnetic field.

The polarization of the dilepton system is analyzed in the rest frame of the virtual photon. We adopt the helicity frame, in which the quantization axis ($z'$-axis) is aligned with the three-momentum of the virtual photon, as illustrated in the right panel of Fig.~\ref{fig:illustration}. In this frame, the lepton and antilepton are emitted back-to-back, and their four-momenta take the form
\begin{equation}
\label{eq:leptonmomentumHX}
\begin{aligned}
l_{1}^\mu &= \left(\frac{M}{2}, |\mathbf{l}| \sin\theta_\ell \cos\phi_\ell, |\mathbf{l}| \sin\theta_\ell \sin\phi_\ell, |\mathbf{l}| \cos\theta_\ell \right)    
\\
l_{2}^\mu &= \left(\frac{M}{2}, -|\mathbf{l}| \sin\theta_\ell \cos\phi_\ell, -|\mathbf{l}| \sin\theta_\ell \sin\phi_\ell, -|\mathbf{l}| \cos\theta_\ell \right),
\end{aligned}
\end{equation}  
where $M=\sqrt{\omega^{2}-\mathbf{q}^{2}}$ is the invariant mass of the dilepton pair, and $|\mathbf{l}| = \sqrt{M^2/4 - m_\ell^2}$ is the magnitude of the lepton three-momentum. The angles $\Omega_\ell = (\theta_\ell, \phi_\ell)$ denote the polar and azimuthal angles of the lepton momentum relative to the quantization axis. From Eq.~\eqref{eq:leptonmomentumHX}, $L_{\mu\nu}$ becomes a function of $\theta_\ell$ and $\phi_\ell$, and the angular distribution follows from its contraction with $\rho^{\mu\nu}$.

The conventional angular distribution of the leptons is expressed as~\cite{Seck:2023oyt}
\begin{equation}
\label{eq:dR2}
\begin{aligned}
\frac{dR_{l^{+}l^{-}}}{d^4 q d\Omega_\ell \, d^4 x}= & \mathcal{N}\left(1+\lambda_\theta \cos ^2 \theta_{\ell}\right. \\
& +\lambda_\phi \sin ^2 \theta_{\ell} \cos 2 \phi_{\ell}+\lambda_{\theta \phi} \sin 2 \theta_{\ell} \cos \phi_{\ell} \\
& \left.+\lambda_\phi^{\perp} \sin ^2 \theta_{\ell} \sin 2 \phi_{\ell}+\lambda_{\theta \phi}^{\perp} \sin 2 \theta_{\ell} \sin \phi_{\ell}\right),
\end{aligned}
\end{equation}
where $\mathcal{N}$ is a normalization factor, and $\lambda_\theta$, $\lambda_\phi$, $\lambda_{\theta\phi}$, $\lambda_{\phi}^{\perp}$, and $\lambda_{\theta\phi}^{\perp}$ are the anisotropy coefficients. These coefficients are intrinsically linked to the polarization state of the virtual photon.

As an illustrative example, consider the decay $\gamma^{*} \rightarrow l^{+}l^{-}$. If the virtual photon is produced in a pure $|1, 0\rangle$ angular-momentum state, its angular distribution is proportional to $1 - \cos^2\theta_{\ell}$. In general, the virtual photon is produced as a statistical mixture of spin states, described by the spin density matrix $\rho_{\lambda \lambda^{\prime}}^{\gamma^{*}}$. The anisotropy coefficients can then be extracted via
\begin{equation}
\begin{array}{r}
\lambda_\theta=\frac{\rho_{-1,-1}^{\gamma^{*}}+\rho_{1,1}^{\gamma^{*}}-2 \rho_{0,0}^{\gamma^{*}}}{\rho_{-1,-1}^{\gamma^{*}}+\rho_{1,1}^{\gamma^{*}}+2 \rho_{0,0}^{\gamma^{*}}}, \\
\lambda_{\theta \phi}=\sqrt{2} \frac{\operatorname{Re}\left(\rho_{0,1}^{\gamma^{*}}-\rho_{-1,0}^{\gamma^{*}}\right)}{\rho_{-1,-1}^{\gamma^{*}}+\rho_{1,1}^{\gamma^{*}}+2 \rho_{0,0}^{\gamma^{*}}}, \\
\lambda_\phi=2 \frac{\operatorname{Re}\left(\rho_{-1,1}^{\gamma^{*}}\right)}{\rho_{-1,-1}^{\gamma^{*}}+\rho_{1,1}^{\gamma^{*}}+2 \rho_{0,0}^{\gamma^{*}}}.
\end{array}
\end{equation}

The primary objective of this work is to investigate how rotation and magnetic fields modify these anisotropy coefficients. A comparison of Eqs.~\eqref{eq:dR1} and~\eqref{eq:dR2} reveals that all relevant information is contained in the spectral function. The spin density matrix is given by~\cite{Speranza:2016tcg}
\begin{equation}
\label{eq:spindensity}
\rho_{\lambda \lambda^{\prime}}^{\gamma^{*}}=\epsilon^{\mu *}(\lambda) W_{\mu \nu} \epsilon^\nu\left(\lambda^{\prime}\right),
\end{equation}
where the tensor $W^{\mu\nu}(q)$ is related to the spectral function $\rho^{\mu\nu}(q)$ through Maxwell's equations~\cite{McLerran:1984ay,Weldon:1990iw}:
\begin{equation}
e^2 W^{\mu\nu}(q) = 2\pi \left( q^2 g^{\mu\alpha} - q^\mu q^\alpha \right) \rho_{\alpha\beta}(-q) \left( q^2 g^{\beta\nu} - q^\beta q^\nu \right).
\end{equation}
In Eq.~\eqref{eq:spindensity}, $\epsilon^\mu(\lambda)$ are the polarization vectors for helicity $\lambda=0,\pm 1$. In the helicity frame, they are explicitly given by
\begin{equation}
\begin{aligned}
\epsilon^\mu(-1) &= \frac{1}{\sqrt{2}} (0, 1, -i, 0),\\
\epsilon^\mu(0) &= (0, 0, 0, 1),\\
\epsilon^\mu(+1) &= -\frac{1}{\sqrt{2}} (0, 1, i, 0). 
\end{aligned}
\end{equation}
Consequently, the spin density matrix elements $\rho_{\lambda \lambda^{\prime}}^{\gamma^{*}}$ can be fully determined from the spectral function $\rho^{\mu\nu}$. In the presence of a rotating or magnetized medium, we derive explicit expressions for both the diagonal and off-diagonal components of $\rho^{\mu\nu}$, which will be discussed in detail in the subsequent sections.

\section{Dilepton polarization under the rotation}
\subsection{Spectral function under a rotating medium}
\label{sec:vorticity}
Quark matter exhibits unique properties under extreme conditions. Vorticity and magnetic fields can influence the motion of quarks and antiquarks in the QCD medium. In a QCD medium, rotational effects can be equivalently described by a fictitious gravitational field \(g^{\mu\nu}\) induced by a comoving rotating reference frame. From the metric \(g^{\mu\nu}\) and the associated vielbein \(e_{a}\,^{\mu}\), one derives both the affine connection \(G^{\sigma}{}_{\mu\nu}\) and the spin connection
\[
\Gamma_{ab\mu}=\eta_{ac}\left(e^{c}{}_{\sigma} G^{\sigma}{}_{\mu\nu}e_{b}{}^{\nu}-e_{b}{}^{\nu}\partial_{\mu}e^{c}{}_{\nu}\right).
\]
The corresponding quantum dynamics are governed by the modified Dirac equation~\cite{Chen:2015hfc,Jiang:2016wvv,Ambrus:2015lfr}
\begin{equation}
\label{eq:dirac1}
    i[\bar{\gamma}^{\mu}(\partial_{\mu}+\Gamma_{\mu})+M_{f}]\psi=0\,,
\end{equation}
where \(M_{f}\) is the quark mass for a given flavor, and \(\Gamma_{\mu}=\frac{1}{8}[\gamma^{a},\gamma^{b}]\Gamma_{ab\mu}\) is the spinor connection. Here, \(\gamma^{a}\) and \(\bar{\gamma}^{\mu}\) are Dirac matrices corresponding to the flat metric \(\eta^{ab}\) and the curved metric \(g^{\mu\nu}\), respectively. The Clifford algebra \(\{\bar{\gamma}^{\mu},\bar{\gamma}^{\nu}\}=g^{\mu\nu}\) requires \(\bar{\gamma}^{\mu}=e_{a}\,^{\mu}\gamma^{a}\).

We consider a rigid rotation perpendicular to the reaction plane, as is customary in the literature~\cite{Chen:2015hfc,Jiang:2016wvv,Ambrus:2023bid}. In this case, the metric takes the form~\cite{Chen:2015hfc}
\begin{equation}
g_{\mu \nu}=\left(\begin{array}{cccc}
1-\left(x^2+y^2\right) \Omega^2 & y \Omega & -x \Omega & 0 \\
y \Omega & -1 & 0 & 0 \\
-x \Omega & 0 & -1 & 0 \\
0 & 0 & 0 & -1
\end{array}\right).
\end{equation}
Consequently, the Dirac equation in a rotating medium simplifies to~\cite{Jiang:2016wvv}
\begin{equation}
\label{eq:dirac2}
    i[\gamma^{\mu}\partial_{\mu}+\Omega\cdot J_{z}+M_{f}]\psi=0\, , 
\end{equation}
where \(J_{z}=L_{z}+S_{z}\) is the third component of the total angular momentum operator. The eigenstates of Eq.~\eqref{eq:dirac2} are given in Ref.~\cite{Jiang:2016wvv}, with energy eigenvalues
\[
E_{p}=\pm\sqrt{p^2+M_{f}^2}-\left(n+\frac{1}{2}\right)\Omega,
\]
where \(n=0,\pm 1,\dots\) is the angular-momentum quantum number. In the rotating frame, the quark propagator is modified by the angular velocity \(\Omega\) under the rigid-rotation approximation~\cite{Castano-Yepes:2025zae, Ayala:2021osy, Vilenkin:1980zv}:
\begin{equation}
\label{eq:propagator-rot}
    S_{\Omega}(p)=\mathcal{P}^{+} \frac{\slashed{p}_{+}+ M_{f}}{p_{+}^{2} - M_{f}^{2} + \mathrm{i}\epsilon} + \mathcal{P}^{-} \frac{\slashed{p}_{-} + M_{f}}{p_{-}^2-M_{f}^{2}+ \mathrm{i}\epsilon},
\end{equation}
where \(\mathcal{P}^{\pm}=1\pm \gamma^{1}\gamma^{2}\) are projection operators, and the quark momenta are shifted as \(p^{\mu}_{\pm}=(p^{0}\pm\frac{\Omega}{2},\mathbf{p})\). The one-loop polarization function under rotation is given by
\begin{equation}
\label{eq:pol}
    \Pi_{\Omega}^{\mu\nu}(q)=-iN_{c}N_{f}\int\frac{\mathrm{d}^{4}p}{(2\pi)^4}\mathrm{Tr}[\gamma^{\mu}S_{\Omega}(p+q)\gamma^{\nu}S_{\Omega}(p)],
\end{equation}
with \(N_{c}=3\) and \(N_{f}=2\) denoting the numbers of colors and flavors, respectively, in this study. The zero-temperature polarization function \(\Pi_{\Omega}^{\mu\nu}(q)\) is extended to finite temperature \(\Pi_{\Omega}^{\mu\nu}(\omega, \mathbf{q})\) via the standard Matsubara frequency summation, where \(\omega\) is the photon energy.

The spectral function \(\rho^{\mu\nu}_{\Omega}(\omega, \mathbf{q})\) is related to the imaginary part of the photon polarization tensor by
\begin{equation}
    \rho^{\mu\nu}_{\Omega}(\omega, \mathbf{q})=-\frac{1}{\pi}\frac{e^{\beta\omega}}{1-e^{-\beta\omega}}\frac{1}{q^{4}}\mathrm{Im}\Pi_{\Omega}^{\mu\nu}(\omega, \mathbf{q}).
\end{equation}
Substituting Eq.~\eqref{eq:propagator-rot} into Eq.~\eqref{eq:pol} and then extracting the imaginary part of the finite-temperature polarization function yields the required components. The diagonal components of \(\mathrm{Im}\,\Pi_{\Omega}^{\mu\nu}\) are explicitly given by
\begin{equation}
\label{eq:pol-rot1}
\begin{aligned}
\mathrm{Im}\,\Pi^{00}_{\Omega}(\omega,\mathbf{q})
=&\frac{3}{4\pi}\sum_{\eta=\pm 1} \int_{p^{0}_-}^{p^{0}_+} p \, \mathrm{d}p\,\frac{4E_{p}^{2}+M^{2}-4\omega E_{p}}{2 q E_p}\\
&\times\left[ 1 - f\left(E_p - \mu - \frac{\eta\Omega}{2}\right) - f\left(\omega - E_p + \mu - \frac{\eta\Omega}{2}\right) \right],
\end{aligned}
\end{equation}
\begin{equation}
\label{eq:pol-rot2}
\begin{aligned}
\mathrm{Im}\,\Pi^{ii}_{\Omega}(\omega,\mathbf{q})
=& -\frac{3}{4\pi}\sum_{\eta=\pm 1} \int_{p^{i}_{-}}^{p^{i}_{+}} \frac{p \, \mathrm{d}p}{q E_p}  \Bigg[ E_p(\omega - E_p - \eta\Omega(1-\delta^{3i}))+\left(1-3n_{i}^2\right)
 \cos^2\theta^{i}_{\eta} \\
 &+p^2 n_{i}^2 +q(1-2n_{i}^2)
\cos\theta^{i}_{\eta}
 + M_f^2 \Bigg] \\
&\times \left[ 1 - f\left(E_p - \mu - \frac{\eta\Omega}{2}\right) - f\left(\omega - E_p + \mu - \frac{\eta\Omega}{2}\right) \right],
\end{aligned}
\end{equation}
where the Latin index \(i=1,2,3\) labels the spatial components. In Eq.~\eqref{eq:pol-rot1} and Eq.~\eqref{eq:pol-rot2}, \(E_p=\sqrt{|\mathbf{p}|^2+M_f^2}\) is the quark energy, and \(q=|\mathbf{q}|\) is the magnitude of the photon three-momentum. The components \(n_{i}=q_{i}/q\) indicate the direction of the virtual photon. The Fermi–Dirac distribution function is \(f(x) = (e^{\beta x}+1)^{-1}\), and the sum over \(\eta = \pm 1\) accounts for the two possible quark-spin orientations relative to the rotation axis.

The integration limits for the quark momentum differ among the various components of the polarization tensor. For the longitudinal component, the upper and lower limits \(p^{0}_{\pm}\) and \(p^{3}_{\pm}\) are given by
\begin{equation}
p^{0,3}_{\pm} = \pm\frac{q}{2} 
+ \frac{\omega}{2}\sqrt{1-\frac{4M_f^2}{\omega^2-q^2}}.
\end{equation}
The transverse components exhibit a more intricate structure due to the rotation-induced anisotropy. Their integration limits \(p^{1}_{\pm}\) and \(p^{2}_{\pm}\) read
\begin{equation}
p_{\pm}^{1,2} = \pm\frac{q}{2} 
+ \frac{\omega+\eta\Omega}{2}\sqrt{1-\frac{4M_f^2}{(\omega+\eta\Omega)^2-q^2}}.
\end{equation}
This \(\Omega\)-dependent deformation of the integration domain reflects the breaking of spherical symmetry in the rotating frame. In Eq.~\eqref{eq:pol-rot1} and Eq.~\eqref{eq:pol-rot2}, \(\cos\theta^{i}_{\eta}\) denotes the angle between the quark momentum \(\mathbf{p}\) and the virtual-photon momentum \(\mathbf{q}\), with the explicit expressions
\begin{equation}
    \cos\theta^{1,2}_{\eta}=\frac{(\omega-\eta\Omega)^2+2(\omega-\eta\Omega)\sqrt{p^2+M_f^2}-q^2}{2q}
\end{equation}
and
\begin{equation}
    \cos\theta^{3}_{\eta}=\frac{\omega^2+2\omega\sqrt{p^2+M_f^2}-q^2}{2q}.
\end{equation}

In summary, the angular velocity enters both the distribution functions (via an effective chemical potential \(\pm\eta\Omega/2\)) and the kinematic boundaries of the phase space, thereby inducing a characteristic polarization pattern in the resulting dilepton spectral function.

\subsection{Numerical Result under the Rotation}
\label{sec:rotating}
\begin{figure}[t]
    \centering
    \includegraphics[width=0.6\linewidth]{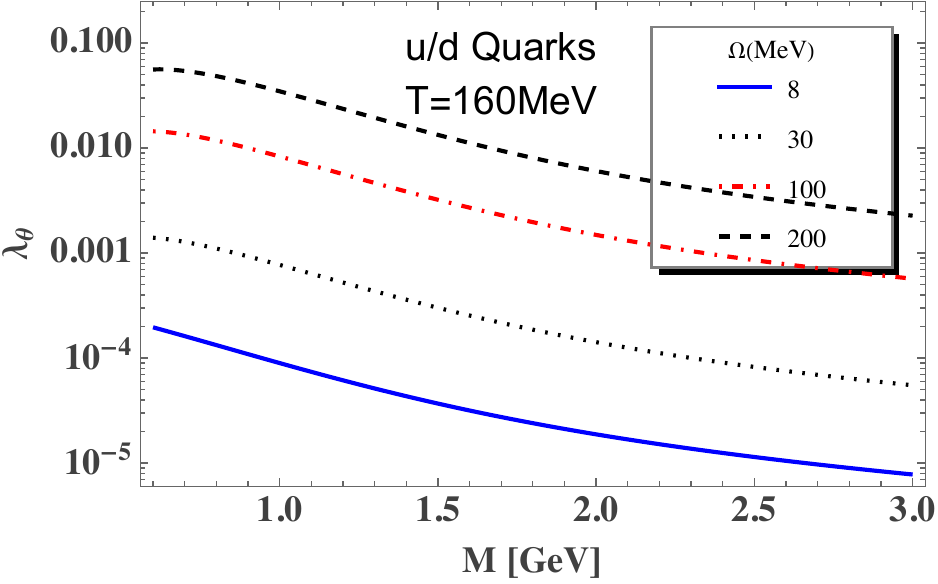}
    \caption{Anisotropic coefficients $\lambda_{\theta}$ as a function of invariant mass $M$ with different angular velocity $\Omega$. The solid, dotted, dot-dashed, and dashed lines stand for $\Omega=8,  30, 100, 200$ MeV, respectively. The virtual photon momentum $\mathbf{q}$ is zero.}
    \label{fig:lambda-theta-rot-log}
\end{figure}
\begin{figure}[t]
    \centering
    \includegraphics[width=0.6\linewidth]{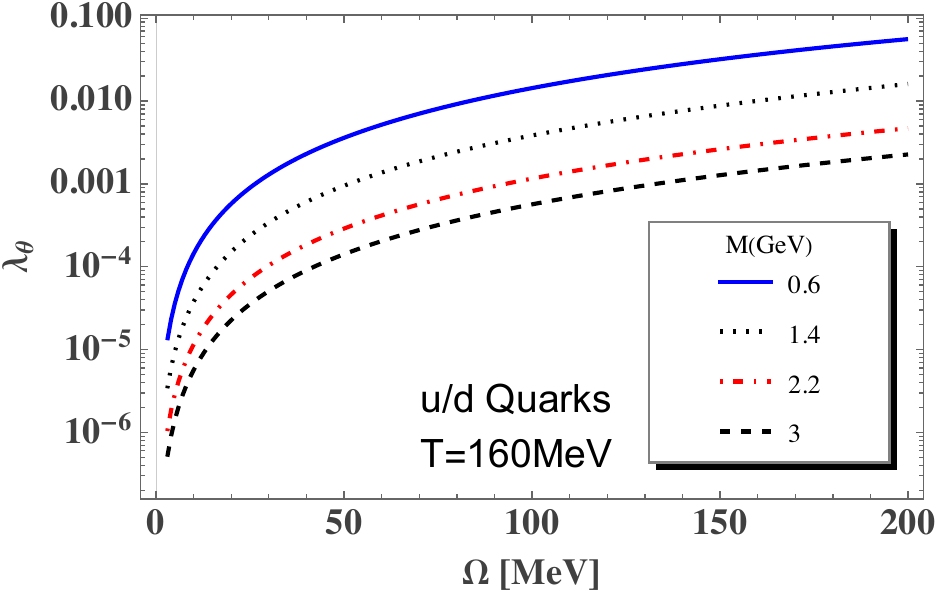}
    \caption{Anisotropic coefficients $\lambda_{\theta}$ as a function of angular velocity $\Omega$ with different invariant mass $M$. The solid, dotted, dot-dashed, and dashed lines stand for $M=0.6, 1.4, 2.2, 3$ GeV, respectively. The virtual photon momentum $\mathbf{q}$ is zero.}
    \label{fig:fig:lambda-theta-rot-log2}
\end{figure}

In the numerical simulations, we take the common quark masses \(M_{u}=M_{d}=5\ \text{MeV}\) and the temperature \(T=160\ \text{MeV}\). In this section, we investigate dilepton polarization under rotation, with the angular velocity of the system chosen as \(\Omega=8,\ 30,\ 100,\ 200\ \text{MeV}\). The averaged global vorticity is about \(8\ \text{MeV}\)~\cite{STAR:2017ckg}, while the strongest global vorticity is estimated to reach approximately \(30\ \text{MeV}\). Furthermore, simulations based on the AMPT model indicate that local vorticity in certain regions can reach values of \(100\)–\(200\ \text{MeV}\)~\cite{Jiang:2016woz}.

Figure~\ref{fig:lambda-theta-rot-log} presents the spin polarization of dileptons (virtual photons) as a function of the invariant mass \(M\) for different magnitudes of the angular velocity \(\Omega\). In a rotating QCD medium, a positive anisotropy coefficient \(\lambda_{\theta}>0\) indicates predominant transverse polarization. As the invariant mass increases, \(\lambda_{\theta}\) decreases significantly and monotonically. For \(\Omega=200\ \text{MeV}\), corresponding to the strongest attainable local vorticity, the dashed line shows that the dilepton polarization is appreciable, reaching \(\lambda_{\theta}=0.056\) at \(M=0.6\ \text{GeV}\). In the global vortical field, where the angular velocity is effectively averaged and thus reduced in magnitude, the polarization signal becomes less pronounced. The solid and dotted lines represent the results for \(\Omega=8\ \text{MeV}\) and \(\Omega=30\ \text{MeV}\), corresponding to the strongest global vortex and the event-averaged scenario, respectively. In these cases, the magnitudes of \(\lambda_{\theta}\) are of order \(10^{-4}\) and \(10^{-3}\), respectively.

Figure~\ref{fig:fig:lambda-theta-rot-log2} displays the order-of-magnitude variation of the dilepton polarization for angular velocities ranging from \(3\ \text{MeV}\) to \(200\ \text{MeV}\). The anisotropy coefficient increases monotonically with increasing \(\Omega\), and exhibits a strong sensitivity to the invariant mass. In the limit \(M \to 0\), the virtual photon approaches a state of complete transverse polarization, resembling that of a real photon.

In a rotating QGP medium, quarks and antiquarks become polarized through spin–orbit coupling. The vorticity field acts symmetrically on particles and antiparticles — that is, it depends only on the coupling between the spin–orbital angular momentum and the vorticity, and not on the sign of the electric charge. This behavior stands in sharp contrast to that of a magnetic field, which exerts opposite forces on particles and antiparticles. Consequently, in a rotating medium, the virtual photon is preferentially polarized into the \(|1,+1\rangle\) state, corresponding to transverse polarization.

\begin{figure}[htbp] 
\centering
\includegraphics[width=0.5\linewidth]{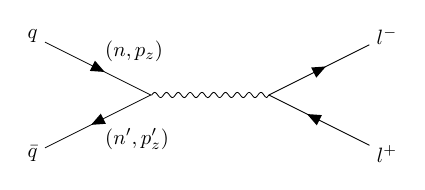}
\caption{The Feynman diagram of quark and anti-quark annihilation in the magnetized medium. The transverse momentum is quantized.}
\label{fig:diagram}
\end{figure}

\section{Dilepton polarization under the magnetic field}
\subsection{Spectral function under the magnetic field}
\label{sec:magnetic-field}
In a magnetized medium, the momenta of quarks and antiquarks are modified, and the magnetic field also aligns the quark spins. The corresponding equation of motion is the Dirac equation
\begin{equation}
    i[\gamma^{\mu}(\partial_{\mu}+q_{f}A_{\mu})+M_{f}]\psi=0\,,
\end{equation}
where \(q_{f}\) is the charge of a quark of a given flavor, and \(A_{\mu}\) is the electromagnetic four-potential, which can be interpreted as an affine connection on a fiber bundle. Numerical simulations suggest that the magnetic field is oriented perpendicular to the reaction plane. We therefore work in the symmetric gauge \(A_{\mu}=(0,-y B/2, xB/2, 0)\), in which the Landau-level wave functions are solved. The dispersion relation yields the discrete energy eigenvalues
\[
E^{2}_{f,n,p_{z}}=p_{z}^{2}+n |q_{f}B|+M_{f}^{2}.
\]
The non-relativistic wave functions are given by
\begin{equation}
     \phi_{n,m}(r, \theta) =N_{n,m} \left( \frac{r}{\ell_{f}} \right)^{|m|} e^{-\frac{r^2}{4\ell_{f}^2}} e^{im\theta} L_n^{|m|}\left( \frac{r^2}{2\ell_{f}^2} \right),
\end{equation}
with
\begin{equation}
     N_{n m} = \sqrt{\frac{n!}{\ell_f^2 \, 2^{|m|} \, (n + |m|)!}},
\end{equation}
where \(L_{n}^{\alpha}\) are the associated Laguerre polynomials, and \(\ell_{f}=1/\sqrt{|q_{f}B|}\) is the magnetic length for a quark of flavor \(f\).

In the lowest Landau level (LLL), quarks and antiquarks are completely spin-polarized: positively charged particles have spin antiparallel to the magnetic field, while negatively charged particles have spin parallel to it. Consequently, in the annihilation process \(q\bar{q}\rightarrow \gamma^{*}\), the virtual photon acquires a net polarization. For higher Landau levels, the polarization becomes partial. As illustrated in Fig.~\ref{fig:diagram}, virtual photons are emitted during transitions of quarks between different Landau levels.

To incorporate contributions from higher Landau levels, we employ the full quark propagator in a magnetized medium~\cite{Miransky:2015ava}:
\begin{equation}
    S_f(p) = i e^{-p_\perp^2 / |q_f B|
} \sum_{n=0}^{\infty} \frac{(-1)^n D_n^f(p)}{\omega_{p}^{2}+p_{z}^{2}-2n|q_{f}B| + i\epsilon},
\end{equation}
where
\begin{equation}
\begin{aligned}
    D_{n}^{f}(\omega_{p}, \mathbf{p}) =& 2\left[ (\omega_{p}+\mu) \gamma^0 - p_{z}\gamma^3 + M_{f} \right]\cdot
\left[ \mathcal{P}^{f}_{+} L_n(2p_{\perp}^{2} \ell_{f}^{2}) - \mathcal{P}^{f}_{-}L_{n-1}(2p_{\perp}^2 \ell_{f}^{2}) \right]\\& + 4(\mathbf{p}_\perp \cdot \mathbf{\gamma}_\perp) L_{n-1}^1(2p_\perp^2 \ell_{f}^{2}).
\end{aligned}
\end{equation}
Here, the projection operators \(\mathcal{P}^{f}_{\pm}=1\pm \mathrm{sign}(q_{f}B)\gamma^{1}\gamma^{2}\) depend on the sign of the quark charge. In the magnetized medium, the one-loop polarization function is defined as
\begin{equation}
    \Pi_{eB}^{\mu\nu}(q)=-i\int\frac{\mathrm{d}^{4}p}{(2\pi)^4}\mathrm{Tr}_{\mathrm{sfc}}[\gamma^{\mu}S_{f}(p+q)\gamma^{\nu}S_{f}(p)],
\end{equation}
where \(\mathrm{Tr}_{\mathrm{sfc}}\) denotes the trace over spin, flavor, and color degrees of freedom. Following the same procedure as in the rotating case, we extract the imaginary part of the polarization function to obtain the longitudinal and transverse conductivities. In the static limit \(\mathbf{q}=0\), these are given by~\cite{Wang:2023fst,Wang:2022jxx}
\begin{equation}
\label{eq:sigma-para}
    \sigma_{\parallel}(\omega) = \sum_{f=u,d} \frac{4\alpha N_c q_f^2}{\omega^2 \ell_f^2} \tanh\left(\frac{|\omega|}{4T}\right) \sum_{n=0}^{n_{\parallel}^{\max}} (2-\delta_{n,0}) \frac{M_{n,f}^2 \theta\left(\omega^2 - 4M_{n,f}^2\right)}{\sqrt{\omega^2 - 4M_{n,f}^2}},
\end{equation}
\begin{equation}
\label{eq:sigma-perp}
\begin{aligned}
\sigma_{\perp}(\omega) =& \sum_{f=u,d} \frac{2\alpha N_c q_f^2 \sinh\left(\frac{\omega}{2T}\right)}{\omega \ell_f^4 \left[\cosh\left(\frac{\omega}{2T}\right) + \cosh\left(\frac{|q_f B|}{T\omega}\right)\right]} \\&\times\sum_{n=1}^{n_{\perp}^{\max}} \frac{\left[2(2n-1) - \omega^2 \ell_f^2\right] \theta\left[(M_{n,f} - M_{n-1,f})^2 - \omega^2\right]}{\sqrt{\left[(M_{n,f} - M_{n-1,f})^2 - \omega^2\right]\left[(M_{n,f} + M_{n-1,f})^2 - \omega^2\right]}} \\
&+ \sum_{f=u,d} \frac{2\alpha N_c q_f^2 \sinh\left(\frac{\omega}{2T}\right)}{\omega \ell_f^4 \left[\cosh\left(\frac{\omega}{2T}\right) + \cosh\left(\frac{|q_f B|}{T\omega}\right)\right]}\\
&\times\sum_{n=1}^{n_{\perp}^{\max}} \frac{\left[\omega^2 \ell_f^2 - 2(2n-1)\right] \theta\left[\omega^2 - (M_{n,f} + M_{n-1,f})^2\right]}{\sqrt{\left[\omega^2 - (M_{n,f} - M_{n-1,f})^2\right]\left[\omega^2 - (M_{n,f} + M_{n-1,f})^2\right]}},
\end{aligned}
\end{equation}
where \(M_{n,f}=\sqrt{2n|q_{f}B|+M_{f}^{2}}\) can be interpreted as an effective mass in the magnetized medium, \(\alpha=1/137\) is the fine-structure constant, and \(\theta(x)\) is the Heaviside step function. The step functions indicate that contributions from higher Landau levels become active only above certain energy thresholds. The maximum Landau-level numbers entering the sums are
\begin{equation}
\begin{aligned}
n_{\max }^{\|}&=\left(\omega^2-4 M_{f}^2\right) /\left(4\left|q_f B\right|\right),\\
n_{\max }^{\perp}&=\left(\left(2\left|q_f B\right|+\omega^2\right)^2-4 M_{f}^2 \omega^2\right) /\left(8\left|q_f B\right| \omega^2\right).
\end{aligned}
\end{equation}

As discussed in Ref.~\cite{Speranza:2017hso}, the polar-angle distribution of the dileptons can be expressed as
\begin{equation}
    \frac{dR_{l^{+}l^{-}}}{d\cos\theta_\ell} \propto \sigma_\perp (1 + \cos^2\theta_\ell) + \sigma_\parallel (1 - \cos^2\theta_\ell).
\end{equation}
Consequently, the anisotropy coefficient \(\lambda_{\theta}\) is given by
\begin{equation}
\begin{aligned}
\lambda_{\theta} &= \frac{\sigma_{\perp}-\sigma_{\parallel}}{\sigma_{\perp}+\sigma_{\parallel}}.
\end{aligned}
\label{eq:lambda-sigma}
\end{equation}

\subsection{Numerical Results under a Magnetic Field}
\label{sec:magnetized}
\begin{figure}[t]
    \centering
    \includegraphics[width=0.6\linewidth]{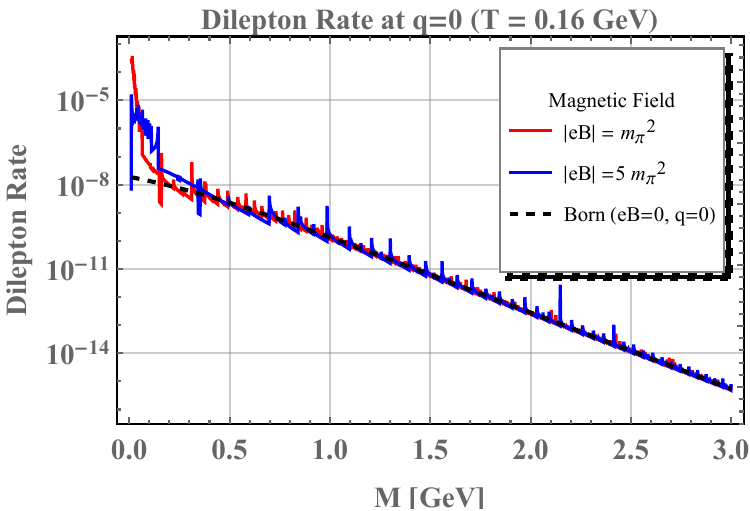}
    \caption{Dilepton production rate in the presence of a magnetic field.}
    \label{fig:dR-EM}
\end{figure}

In this section, we adopt the same quark masses \(M_{u}=M_{d}=5\ \text{MeV}\) and temperature \(T=160\ \text{MeV}\) as in the rotating case. For the numerical simulations, we need to specify the strength of the magnetic field.

In the initial stage of heavy-ion collisions (HICs), the magnetic field can be extremely large. In \(\sqrt{s_{NN}}=200\ \text{GeV}\) Au-Au collisions at RHIC, its magnitude is approximately several \(m_{\pi}^{2}\). At LHC energies, the initial magnetic field can reach \(eB\sim 10m_{\pi}^{2}\) or even higher. However, the magnetic field is primarily generated by the spectator nucleons, which separate rapidly from each other, causing the magnitude of \(eB\) to decrease significantly over time. Regarding the magnetic field strength during the QGP evolution stage, key experimental evidence is lacking, and theoretical understanding remains controversial. For theoretical study, we therefore consider a relatively wide range of values, \(eB\sim 1\)–\(10\,m_{\pi}^{2}\).

Figure~\ref{fig:dR-EM} displays the invariant-mass spectrum of the thermal dilepton production rate for different magnetic field strengths. The black dashed line represents the Born rate at zero magnetic field, which serves as the baseline corresponding to a purely thermal QGP medium. In this case, the quark–antiquark annihilation process dominates the rate, whose magnitude decreases smoothly with increasing invariant mass \(M\). For nonzero magnetic fields, the red and blue solid lines correspond to \(eB=m_{\pi}^{2}\) and \(eB=5m_{\pi}^{2}\), respectively. The curves in Fig.~\ref{fig:dR-EM} show that the magnetic field significantly enhances the dilepton production rate in the low invariant-mass region (\(M \lesssim 0.6\ \text{GeV}\)). In the higher-mass region (\(M \gtrsim 0.6\ \text{GeV}\)), the magnetic-field effect gradually weakens, and the curves tend to merge with or approach the zero-field case. For a more detailed discussion, see Ref.~\cite{Wang:2022jxx}.

\begin{figure*}
    \centering
    \includegraphics[width=0.45\linewidth]{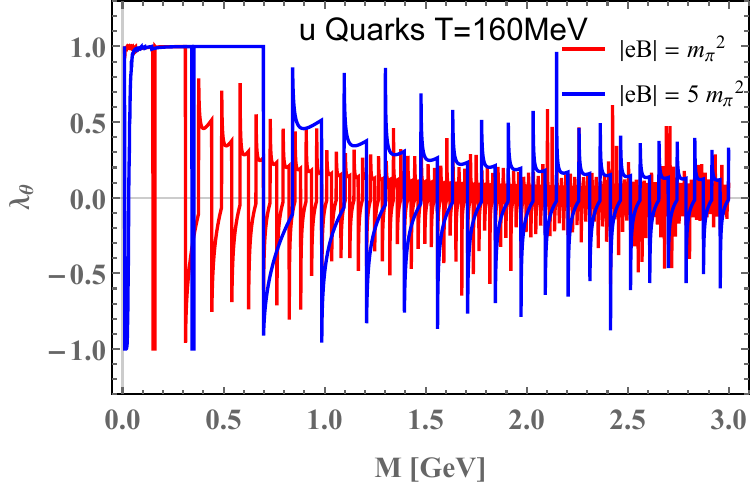}
    \includegraphics[width=0.45\linewidth]{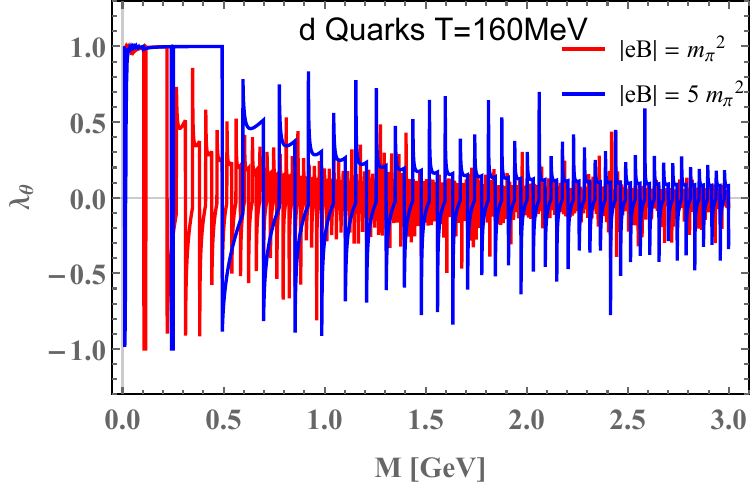}
    \includegraphics[width=0.45\linewidth]{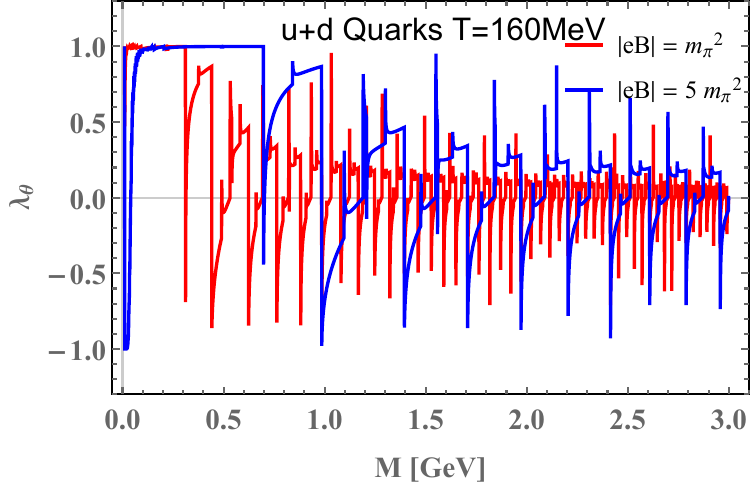}
    \caption{Dilepton polarization under magnetic field. The anisotropic coefficients $\lambda_{\theta}$ vary as a function of invariant mass $M$. The red and blue solid lines stand for $eB=m_{\pi}^{2}$  and $eB=5 m_{\pi}^{2}$ correspondingly.}
    \label{fig:eB-ud-M}
\end{figure*}

\begin{figure*}
    \centering
    \includegraphics[width=0.45\linewidth]{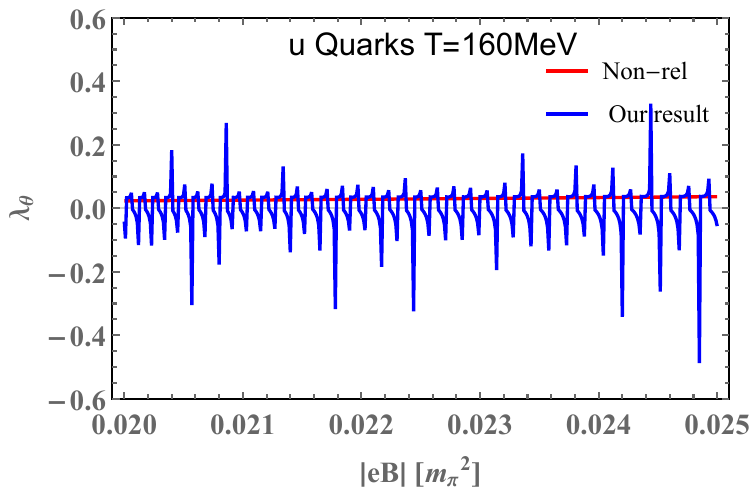}\hspace{0.5cm}
    \includegraphics[width=0.45\linewidth]{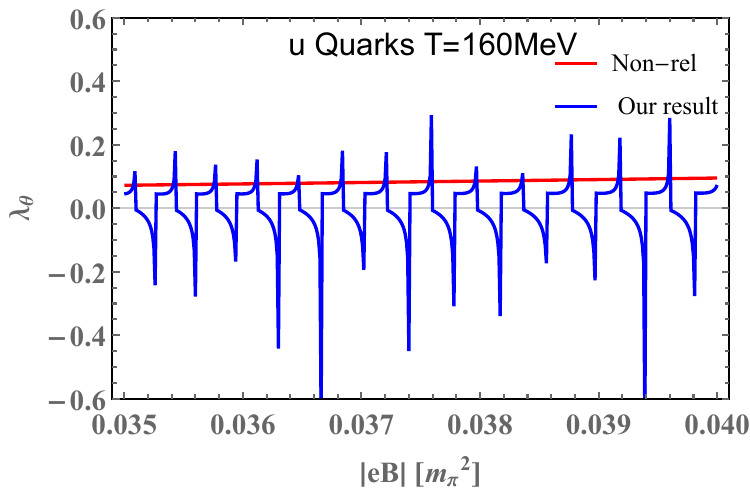}\\
    \vspace{0.8cm}
    \includegraphics[width=0.45\linewidth]{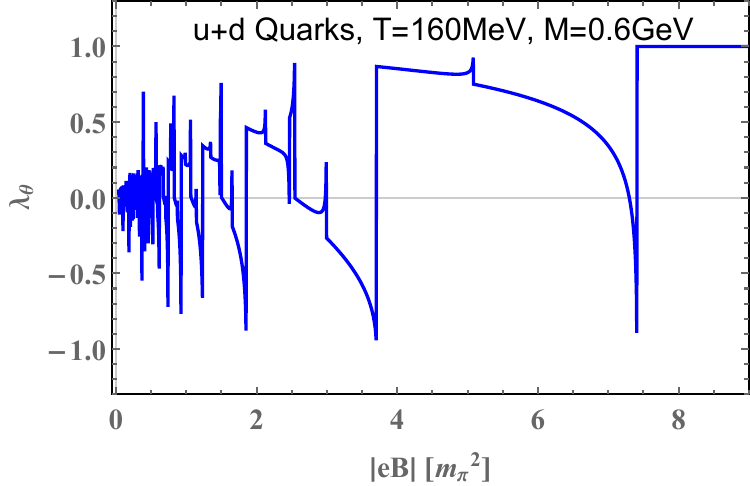}
    \hspace{0.3cm}
    \includegraphics[width=0.45\linewidth]{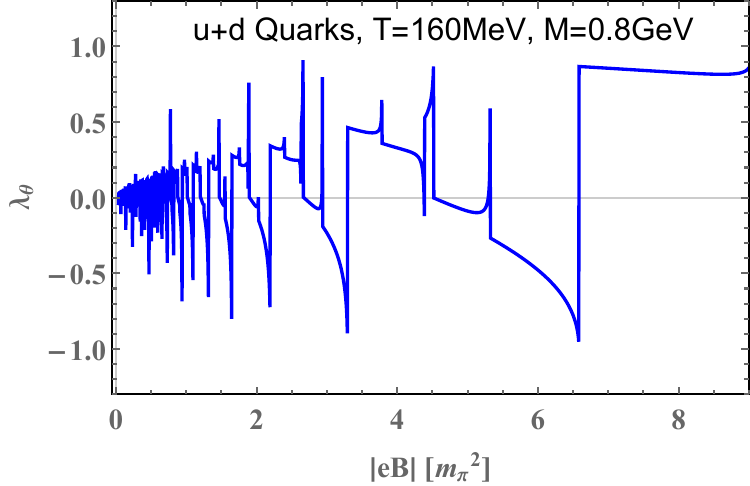}
    \caption{Anisotropic coefficients $\lambda_{\theta}$ as a function of magnetic field $|eB|$.}
    \label{fig:eB-u-eB}
\end{figure*}

Unfortunately, thermal dileptons dominate in the invariant-mass range \(M\sim 0.6\)–\(3\ \text{GeV}\). In this region, dilepton polarization emerges as a more promising probe of the magnetic field. In the absence of a magnetic field, virtual photons are unpolarized, and the anisotropy coefficient \(\lambda_{\theta}\) vanishes. Figures~\ref{fig:eB-ud-M} and~\ref{fig:eB-u-eB} demonstrate that a nonzero magnetic field induces a significant dilepton polarization. In particular, Fig.~\ref{fig:eB-ud-M} presents the invariant-mass spectrum of \(\lambda_{\theta}\), showing that the magnetic-field-induced polarization remains substantial even in the high-mass region from \(0.6\) to \(3\ \text{GeV}\).

To clarify the underlying physics, we first isolate the individual contributions of up and down quarks. In the first panel of Fig.~\ref{fig:eB-ud-M}, the red and blue lines represent \(\lambda_{\theta}\) as a function of \(M\) for \(eB=m_{\pi}^{2}\) and \(eB=5m_{\pi}^{2}\), respectively. The two curves display similar overall behavior, but the Landau levels are more closely spaced in the weaker field. Thus, the blue curve, with more widely separated Landau levels, exhibits a clearer oscillatory structure, and our analysis of these features will primarily focus on the \(eB=5m_{\pi}^{2}\) case.

In general, \(\sigma_{\parallel}\) contributes to longitudinal polarization, while \(\sigma_{\perp}\) contributes to transverse polarization. The oscillatory pattern of the curve arises from the step functions appearing in Eqs.~\eqref{eq:sigma-para} and~\eqref{eq:sigma-perp}. These functions indicate the thresholds at which different Landau levels start to contribute to either the transverse or longitudinal polarization modes.

When \(M<M_{1,u}-M_{0,u}=0.344\ \text{GeV}\) or \(M>M_{1,u}+M_{0,u}=0.354\ \text{GeV}\), complete transverse polarization is induced by the \(n=1\) term in Eq.~\eqref{eq:sigma-perp}. For the \(eB=5m_{\pi}^{2}\) case, \(\lambda_{\theta}\) remains nearly constant at \(\lambda_{\theta}=1\) for \(M<0.697\ \text{GeV}\).

When the invariant mass satisfies \(M_{1,u}-M_{0,u}<M<M_{1,u}+M_{0,u}\) or lies close to \(2M_{0,u}=0.01\ \text{GeV}\), complete longitudinal polarization is induced by the \(n=0\) term in Eq.~\eqref{eq:sigma-para}. As discussed in the previous section, charged particles in the lowest Landau level are fully spin-polarized, and consequently the virtual photon is completely longitudinally polarized in this region.

At \(2M_{1,u}=4\sqrt{5/3}\,m_{\pi}=0.697\ \text{GeV}\), \(\sigma_{\parallel}\) begins to contribute longitudinal polarization, resulting in a downward jump in the curve at this point. At \(M=M_{2,u}+M_{1,u}=0.842\ \text{GeV}\), \(\sigma_{\perp}\) begins to contribute transverse polarization, causing an upward discontinuity. By the same mechanism, transverse and longitudinal contributions alternate, producing an oscillatory behavior in the curve. The oscillation amplitude gradually decreases, as quarks in higher Landau levels are less polarized.

The second panel of Fig.~\ref{fig:eB-ud-M} shows the results for down quarks, which exhibit a similar oscillatory pattern. Owing to the fractional charge \(-1/3\) of the down quark, its oscillatory features are more closely spaced. The third panel presents the combined results including both up and down quarks, where the oscillation "period" is effectively widened. In summary, all panels consistently indicate that the dilepton polarization driven by a strong magnetic field exceeds that driven by a weak field.

Next, we compare our results with those of the non-relativistic quark coalescence model. This model provides a spin density matrix element \(\rho_{00}\) associated with the magnetic field~\cite{Yang:2017sdk}, given explicitly by
\begin{equation}
    \rho_{00}^{\text{Non-rel. coal.}} \approx \frac{1}{3} - \frac{1}{9T^{2}} \frac{q_{f} q_{f'}}{M_{f}M_{f'}} B^2.
\end{equation}
In this expression, the spin density matrix element is a quadratic function of the magnetic field, which evidently corresponds to the classical approximation in the weak-field limit.

Figure~\ref{fig:eB-u-eB} displays the anisotropy coefficient \(\lambda_{\theta}\) as a function of \(|eB|\), with the invariant mass fixed at \(M=0.6\ \text{GeV}\). For clarity of the physical analysis, the first two panels include only up-quark and anti-up-quark contributions, with magnetic-field ranges of \(0.020 - 0.025m_{\pi}^{2}\) and \(0.035 - 0.040m_{\pi}^{2}\), respectively. These ranges are chosen because a magnetic field that is too small would require an impractically large number of Landau levels in the calculation, while a field that is too large would exceed the applicability of the quark coalescence model. The blue lines represent our numerical results for \(\lambda_{\theta}\), while the red lines show the corresponding predictions from the quark coalescence model for comparison.

In the weak-field range \(0.020 - 0.025m_{\pi}^{2}\), our results agree well in magnitude with the classical coalescence model when quantum oscillations are neglected. For fields in the range \(0.035 - 0.040m_{\pi}^{2}\), a noticeable deviation appears, indicating that quantum effects become significantly enhanced and the classical description is no longer applicable. Additionally, as seen from the first and second panels, a stronger magnetic field leads to more widely spaced Landau levels.

In the last two panels of Fig.~\ref{fig:eB-u-eB}, both up- and down-quark contributions are included, and \(\lambda_{\theta}\) is plotted against the magnetic field strength ranging from \(0.05m_{\pi}^{2}\) to \(9m_{\pi}^{2}\). In the strong-field region, the Landau levels are so widely separated that only a single-digit number of levels contributes. A comparison between the curves at invariant masses of \(0.6\ \text{GeV}\) and \(0.8\ \text{GeV}\) shows that the latter behaves like a horizontally stretched version of the former.

\begin{figure}
    \centering
    \includegraphics[width=0.6\linewidth]{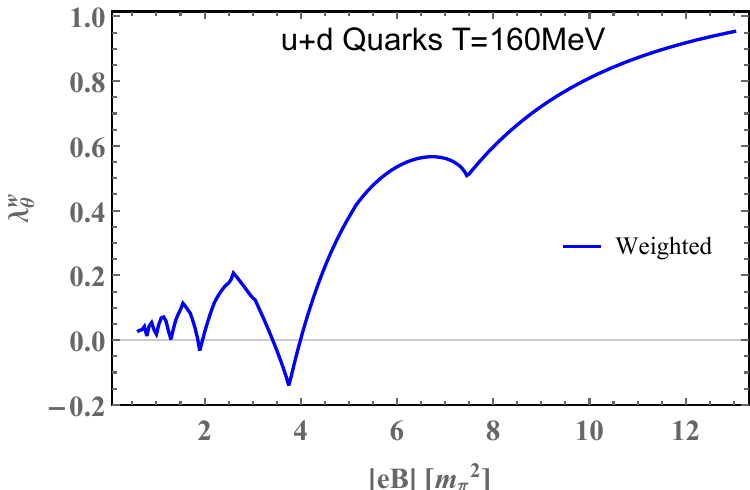}
    \caption{The average $\lambda_{\theta}$ as a function of magnetic field $|eB|$.}
    \label{fig:eB-ud-eB-weighted}
\end{figure}

Finally, we investigate the averaged anisotropy coefficient as a function of \(|eB|\). We integrate \(\lambda_{\theta}(M)\) over the invariant-mass interval from \(0.6\ \text{GeV}\) to \(3\ \text{GeV}\), weighted by the dilepton production rate:
\begin{equation}
    \lambda_{\theta}^{\mathrm{w}}=\frac{1}{R_{l^{+}l^{-}}}\int \frac{\mathrm{d}R_{l^{+}l^{-}}}{\mathrm{d}M}\lambda_{\theta}(M)\,\mathrm{d}M,
\end{equation}
where the normalization factor \(R_{l^{+}l^{-}}\) denotes the total dilepton yield integrated over the mass range \([0.6, 3]\ \text{GeV}\).

The numerical result is shown in Fig.~\ref{fig:eB-ud-eB-weighted}. After the weighted averaging procedure, the anisotropy coefficient becomes considerably smoother, with most of the sharp quantum-transition signatures smeared out. In the third panel of Fig.~\ref{fig:eB-u-eB}, two abrupt jumps are visible at \(|eB|=3.7m_{\pi}^{2}\) and \(|eB|=7.4m_{\pi}^{2}\). In Fig.~\ref{fig:eB-ud-eB-weighted}, however, these discontinuities are softened into mere inflection points, shifted to \(|eB|=3.8m_{\pi}^{2}\) and \(|eB|=7.5m_{\pi}^{2}\), respectively. Since the dilepton production rate falls off exponentially with increasing invariant mass, the yield near \(M=0.6\ \text{GeV}\) carries a disproportionately large weight. Consequently, the locations of the turning points exhibit only minor shifts.

\section{Summary and Outlook}
\label{sec:summary}
In this work, we have systematically investigated the polarization characteristics of dileptons (virtual photons) produced in a hot quark-gluon plasma (QGP) under two extreme conditions: a vorticity field (rotation) and a strong magnetic field. In a rotating medium or a magnetized medium, we computed the photon polarization tensor. The spectral function was decomposed into the longitudinal component and the transverse component.   

In a rotating medium, the anisotropy coefficient $\lambda_{\theta}$ decreases monotonically with increasing invariant mass $M$. For the strongest local vorticity ($\Omega = 200\,\text{MeV}$), $\lambda_{\theta}$ reaches $0.056$ at $M = 0.6\,\text{GeV}$. In contrast, for globally averaged vorticity, the polarization magnitude is significantly smaller, on the order of $10^{-3}$ to $10^{-4}$.
    
In a magnetized medium, for invariant masses smaller than a certain threshold, the dileptons maintain transverse polarization. As higher Landau levels become accessible with increasing $M$, longitudinal and transverse contributions alternate, producing an oscillatory behavior in $\lambda_{\theta}$ with gradually decaying amplitude. After applying a weighted average weighted by the dilepton production rate, the oscillatory features are significantly weakened. In general, the magnetic field leads to transverse polarization of the dileptons.

Based on the results presented in this paper, several directions merit further investigation.
The most intriguing target is the combined effects of rotation and magnetic field. In realistic heavy-ion collisions, both vorticity and magnetic field coexist and may interfere. Extending the present formalism to include both $\Omega$ and $eB$ simultaneously, and exploring their cooperative or competing influences on dilepton polarization, is a natural and important next step.
    
Moreover, non-rigid rotation and inhomogeneous vorticity are also interesting. The rigid rotation approximation can be relaxed to incorporate a radially dependent angular velocity, as suggested by hydrodynamic simulations. This would allow a more realistic comparison with event-by-event fluctuations of the local vorticity.
    
While this work focused primarily on $\lambda_{\theta}$, the full angular distribution involves additional coefficients such as $\lambda_{\phi}$, $\lambda_{\theta\phi}$, $\lambda_{\phi}^{\perp}$, and $\lambda_{\theta\phi}^{\perp}$. A comprehensive analysis of their responses to magnetic fields and vorticity could provide a multi-dimensional probe for discriminating between different medium properties.
    
Finally, with the upcoming high-precision experiments at sPHENIX, ALICE Upgrade, and the future CMC detector, we anticipate that systematic measurements of dilepton polarization will become feasible across a wide range of collision systems and energies. This will offer a unique and powerful tool to constrain the transport properties, initial geometry, and topological structures of the QCD matter created in relativistic heavy-ion collisions.

\acknowledgments
%\begin{acknowledgments}
We wish to acknowledge the support of the National Natural Science Foundation of China (NSFC No. 12505147). This work is also supported by the start-up funding from the Anhui University of Science and Technology(No. 2024yjrc157). X.W. was supported by the Anhui University
of Science and Technology under Grant No. YJ20240001. During the preparation of this manuscript, the authors used DeepSeek-v4-flash for language polishing and proofreading purposes only.
%\end{acknowledgments}
\bibliographystyle{JHEP}
\bibliography{biblio.bib}
\end{document}